\documentclass[14pt,prb ,article]{revtex4-1}
\usepackage{amsmath}  % needed for \tfrac, \bmatrix, etc.
\usepackage{amsfonts} % needed for bold Greek, Fraktur, and blackboard bold
\usepackage{graphicx} % needed for figures
\usepackage{enumitem} % needed to ''close up'' itemizations
\setitemize{noitemsep,topsep=0pt,parsep=0pt,partopsep=0pt}
\usepackage{bm}    % needed for bold Greek letters in math
\usepackage{hyperref}    % needed for hyperlinks to references
\usepackage{extsizes}    % enlarge font size for review
\usepackage[margin=70pt]{geometry}
\begin{document}

\title{The Ubiquitous Lorentz Force}

\author{Clinton L. Lewis}
\affiliation{Division of Science and Mathematics, West Valley College, Saratoga, California (retired)}
\email{CLewis@Swiftwords.com}

\date{\today}

\begin{abstract}
Minimizing the Action integral of a Lagrangian provides the Euler-Lagrange equation of motion in the elegant machinery of Lagrangian Mechanics. However two relations define the divergence of current and energy-momentum, and provide an alternative motivation for the Euler-Lagrange equation without invoking the considerable machinery required for the principle of least Action. The derivation of these two relations proceeds with only local differential operations on the Lagrangian, and without a global integration defining an Action. The two relations connect local continuity equations (as a vanishing divergence) for current and energy-momentum to Lorentz force, symmetry, and the Euler-Lagrange equation. The Euler-Lagrange equation is common to both relations so providing sufficient motivation for acceptance as equations of motion. The essential relationships between the central concepts of energy-momentum, force, current, symmetry and equation of motion provide pedagogically interesting clarity. The student will see that  how these concepts relate to each other as well as the definition of each concept in isolation.
\end{abstract}

\maketitle

\section{Introduction}

Lagrangian Mechanics as applied to classical field theory provides an elegant and general formalism connecting equations of motion, symmetries and the definition of charged currents. The equations of motion for many systems subject to external forces are derivable from specifying a Lagrangian function, then applying the principle of least action\cite{CarrollP159} to the Action integral to derive the Euler-Lagrange equation of motion for the field. If the Action integral has internal symmetries, then charged currents are subsequently derived from the definition of the symmetry.

This paper follows the well-trodden path of Lagrangian mechanics, but \textit{without} the principle of least action. The Lagrangian remains at the center of this ``local'' approach, but no integrals are to be found. The local approach leads to the same Euler-Lagrange equation, as it must, but with different motivations yielding interesting insights for the student already familiar with the principle of least action.

The presentation is self-contained for the curious student, and follows a step-by-step derivation of two equations which are fundamental in physics, one which defines the divergence of the energy-momentum tensor and the other defines the divergence of the current associated with a gauge transformation The derivation uses only the chain rule and product rule for derivatives.

\section{The System Description}

The \textbf{stage} for the evolution of fields in our system is a general space-time metric. The metric, although arbitrary, is a fixed background metric so that Einstein equations do not apply. The \textbf{notation} emphasizes covariance, so tensor notation, Einstein summation and covariant derivatives are used throughout. The development, although covariant, is not general relativistic since the system considered here is subject to a force due to an external gauge field, a situation impossible in general relativity where there are no external forces.

A vector-matrix notation is used to avoid a paroxysm of indices. \textbf{Bold type} indicates a matrix with accompanying linear homogeneous gauge transformation properties.

The a-priori objects in our system are defined throughout spacetime:

\begin{itemize}
\item The wave function ${\bm\phi }$ -- a multicomponent field presented as a column matrix,
\item an external potential field and associated gauge transformation,
\item a gauge covariant derivative of the wave function ${D_\mu }{\bm\phi }$ -- a column matrix,
\item a scalar Lagrangian which is a function of the wave function and its gauge covariant derivative, and the metric of space-time,
\item an external gravitational field in the form of a background metric (not General Relativity).
\end{itemize}

The \textbf{equation of motion} of the multicomponent classical field   (wave function) is defined by the covariant Euler-Lagrange equation as derived from the Lagrangian. The system evolves under the influence of a gravitational field and an external electromagnetic field.

\section{The form of the Lagrangian implies a differential relation}
Consider a general Lagrangian ${L}$ for a system in spacetime with coordinates ${x^\mu }$  and with a multicomponent wave function ${\bm\phi }\left( x \right)$  represented as a column matrix (\textbf{bold type} indicating a matrix) with each complex-valued matrix element a function of space-time coordinates, and a given metric ${g^{\mu \nu }}\left( x \right)$. The Lagrangian is dependent upon the space-time coordinates only through the wave function and fixed background metric, so that there is no coordinate ''${x}$'' appearing explicitly in the Lagrangian.
\begin{equation}
\label{Lagrangian}
L = L\left( {{\bm\phi },{D_\mu }{\bm\phi },{g^{\mu \nu }}} \right)
\end{equation}
The symbol ''${D_\mu }$'' indicates a \textit{gauge covariant} partial derivative with respect to coordinate $x^\mu$ which, like all derivatives, satisfies the product rule (Leibnitz rule). A detailed definition is provided later, but for now simply note that it does not commute as does the ordinary partial derivative with respect to coordinates ${\partial }/{{\partial {x^\mu }}}$, or in short form  ${\partial _\mu }$.

The covariant derivative of the metric is defined to vanish, the \textit{metricity} condition.\cite{BlagoP20}
\begin{equation}
\label{metricity} {D_\lambda }{g^{\mu \nu }}\left( x \right) = {\nabla _\lambda }{g^{\mu \nu }}\left( x \right) = 0
\end{equation}
The gauge covariant derivative $D_\mu$ reverts to the coordinate covariant derivative $\nabla _\lambda $ when applied to gauge invariant objects such as the metric.
 
 Apply the gauge covariant derivative to the scalar Lagrangian, and use the chain rule. It is difficult to motivate taking this derivative, except that, upon completion of a few steps, three major features of Lagrangian mechanics are connected in a single differential relation.
\begin{equation}
\label{chainR}
{D_\mu }L = \frac{{\partial L}}{{\partial {\bm\phi }}}{D_\mu }{\bm\phi } + \frac{{\partial L}}{{\partial \left( {{D_\nu }{\bm\phi }} \right)}}{D_\mu }\left( {{D_\nu }{\bm\phi }} \right) + \frac{{\partial L}}{{\partial \left( {{g^{\lambda \nu }}} \right)}}{D_\mu }{g^{\lambda \nu }}
\end{equation}
\begin{quote}\small
Note:  the partial with respect to the column matrix $\bm\phi $  is to be taken with respect to its components so that  $\partial L/\partial {\bm\phi }$ is a row matrix and ${D_\mu }{\bm\phi }$  is a column matrix. The matrix multiplications in Eq.~(\ref{chainR}) provide the summations required in the chain rule.
\end{quote}
The last term vanishes by the metricity condition above. Add and subtract the second order derivative, but with swapped indices.
\begin{equation}
\begin{gathered}
  {D_\mu }L = \frac{{\partial L}}{{\partial {\phi }}}{D_\mu }{\phi } + \frac{{\partial L}}{{\partial {D_\nu }{\phi }}}{D_\mu }{D_\nu }{\phi } \\ 
  { - }\frac{{\partial L}}{{\partial {D_\nu }{\phi }}}{D_\nu }{D_\mu }{\phi } + \frac{{\partial L}}{{\partial {D_\nu }{\phi }}}{D_\nu }{D_\mu }{\phi } \\ 
\end{gathered} 
\end{equation}
Use a commutator bracket,
\begin{equation}
\label{useComm}
{D_\mu }L = \frac{{\partial L}}{{\partial {\bm\phi }}}{D_\mu }{\bm\phi } + \frac{{\partial L}}{{\partial {D_\nu }{\bm\phi }}}\left[ {{D_\mu },{D_\nu }} \right]{\bm\phi } + \frac{{\partial L}}{{\partial {D_\nu }{\bm\phi }}}{D_\nu }{D_\mu }{\bm\phi }
\end{equation}
and apply the product rule on the last term on the right,
\begin{equation}
\label{prodRule}
\begin{gathered}
  {D_\mu }L = \frac{{\partial L}}{{\partial {\bm\phi }}}{D_\mu }{\bm\phi } + \frac{{\partial L}}{{\partial {D_\nu }{\bm\phi }}}\left[ {{D_\mu },{D_\nu }} \right]{\bm\phi } \\ 
   + {D_\nu }\left( {\left( {\frac{{\partial L}}{{\partial {D_\nu }{\bm\phi }}}} \right){D_\mu }{\bm\phi }} \right){ - }{D_\nu }\left( {\frac{{\partial L}}{{\partial {D_\nu }{\bm\phi }}}} \right){D_\mu }{\bm\phi } \\ 
\end{gathered} \end{equation}
 then rearrange to expose the Euler-Lagrange equation in the square bracket.
\begin{equation}
\label{ELexpose}
\begin{gathered}
  {D_\mu }L = \left[ {\frac{{\partial L}}{{\partial {\bm\phi }}}{ - }{D_\nu }\left( {\frac{{\partial L}}{{\partial {D_\nu }{\bm\phi }}}} \right)} \right]{D_\mu }{\bm\phi  + }\frac{{\partial L}}{{\partial {D_\nu }{\bm\phi }}}\left[ {{D_\mu },{D_\nu }} \right]{\bm\phi } \\ 
  { + }{D_\nu }\left( {\frac{{\partial L}}{{\partial {D_\nu }{\bm\phi }}}{D_\mu }{\bm\phi }} \right) \\ 
\end{gathered}
\end{equation}
Identify the equation of motion for the wave function ${\bm\phi }$ provided by the covariant Euler-Lagrange equation $\Lambda  = 0$ (row matrix),\cite{CLewis,CarrollP160,LandauRefp77_322}
\begin{equation}
\label{ELequ}
\Lambda  = \frac{{\partial L}}{{\partial {\bm\phi }}} - {D_\mu }\left( {\frac{{\partial L}}{{\partial {D_\mu }{\bm\phi }}}} \right) = 0
\end{equation}
\begin{quote}\small
Note:  The many examples of wave equations defined by the Euler-Lagrange equation include electromagnetics, the Klein-Gordon wave equation for spin zero fields and the Dirac wave equation for spin one-half fields. Satisfaction of the Euler-Lagrange equation, $\Lambda  = 0$, defines the term  ``on shell''. Off shell dynamics violate the Euler-Lagrange equation. We will refer to the Euler-Lagrange expression, as well as the equation where the expression is set to zero.
\end{quote}

Substitute the Euler-Lagrange expression $\bm\Lambda$ into Eq.~(\ref{ELexpose}), move terms, and insert the Kronecker delta tensor.
\begin{equation}
\label{EMTexpose}
{D_\nu }\left( {\delta _\mu ^\nu L}  -  {\frac{{\partial L}}{{\partial {D_\nu }{\bm\phi }}}{D_\mu }{\bm\phi }} \right) = \frac{{\partial L}}{{\partial {D_\nu }{\bm\phi }}}\left[ {{D_\mu },{D_\nu }} \right]{\bm\phi } + \Lambda {D_\mu }{\bm\phi }
\end{equation}
Identify the  canonical energy-momentum tensor ${T_{\mu }}^\nu$ (EMT), which is the covariant generalization of energy and momentum in space time,\cite{canonicalEMT,LandauRef2,MichioP27e165}
\begin{equation}
\label{EMT}
{T_{\mu }}^\nu  = L\delta _\mu ^\nu { - }\frac{{\partial L}}{{\partial {D_\nu }{\bm\phi }}}{D_\mu }{\bm\phi }
\end{equation}
then substitute the EMT into Eq.~(\ref{EMTexpose}) to find the divergence of the EMT. The gauge covariant derivative $D _\nu $ becomes the ordinary covariant derivative $\nabla _\nu $ when applied to the gauge-invariant EMT.
\begin{equation}
\label{theIdent}
\boxed{{\nabla _\nu }\left( {{T_\mu}^\nu } \right) = \left( {\frac{{\partial L}}{{\partial {D_\nu }{\bm\phi }}}} \right)\left[ {{D_\mu },{D_\nu }} \right]{\bm\phi } + \bm\Lambda {D_\mu }{\bm\phi }}
\end{equation}
This is the three-part differential relation we are seeking.

The relation between 1) the divergence of the EMT, 2) non-vanishing commutator of the gauge covariant derivative, and 3) the Euler-Lagrange equation of motion is made apparent, and is independent of the principle of least action, and independent of any space-time symmetries of the Lagrangian. The differential relation is a consequence of 1) the form of the Lagrangian, 2) the chain rule and the product rule.

The non-commuting derivative remains to be defined as applied to the wave function, and it remains to associate the commutator with a force as will be shown later. However, as a side note, the differential relation can be applied to any function of the form in Eq.~(\ref{Lagrangian}) which supports a derivative. The Lagrangian is \textit{not} required to be a scalar, nor the indices \textit{tensor} indices.

External forces are modeled in the non-commuting gauge covariant derivative. In order to complete the physical interpretation of Eq.~(\ref{theIdent}), we will find the commutation to be related to a Lorentz force acting on a charged current.

In an appendix, the local procedure for deriving the Euler-Lagrange equation is applied to a Lagrangian for the Klein-Gordon equation for a complex scalar wave function.

\section{The gauge transformation defines current}
Connecting the commutator of the gauge covariant derivative in the differential relation  Eq.~(\ref{theIdent}) to the Lorentz force occupies the remainder of this paper. Gauge theory provides the necessary connections to the Lorentz force on charged currents, so, for brevity, we confine ourselves to the \textit{abelian} Lie algebra of electromagnetics. Perhaps a slight extension is that the wave function is a column matrix of complex values rather than a single complex value. However, a slight extension is that the wave function is a column matrix of complex values rather than a single complex value, but still the gauge transformation is the single real parameter $U\left( 1 \right)$  transformation of electromagnetics and quantum mechanics.

The citations found in subsequent paragraphs refer to more advanced texts which include non-abelian and multi-parameter gauge transformations. The similarity of notation and concepts used here may allow easier comprehension of these more advanced texts.

In this section, we study how the wave function, its derivative and the Lagrangian respond to an infinitesimal gauge transformation, and conspire to define a current, which enters into the definition of the Lorentz force.

Assume the wave function ${\bm\phi }$ (column matrix of complex values) responds to an infinitesimal gauge transformation parameterized with a single real-valued number ${\theta\left( x \right)}$ which is also function of position. Consider an arbitrary infinitesimal variation of the parameter $\delta {\theta}\left( x \right)$, and \textit{define} the wave function $\delta{\bm\phi }$ response to the infinitesimal gauge transformation as,\cite{Wein2}
\begin{equation}
\label{gtrans} \delta {\bm\phi  = }i{\bm\phi }\delta \theta 
\end{equation}
The insertion of the imaginary ``$i$'' is useful later for interacting with gauge fields which are defined as real-valued according to convention. The infinitesimal gauge transformation may also be presented in the form of a derivative.
\begin{equation}
\label{gaugeTransDeriv}
\delta {\bm\phi /}\delta \theta { = }i{\bm\phi }
\end{equation} By construction, require the derivative of the field to have the same response to the variation as the field itself (covariant), so that the derivative is a \textit{gauge covariant} derivative.\cite{DoughtyP455}
\begin{equation}
\label{dgtrans} \delta \left( {{D_\mu }{\bm\phi }} \right){/}\delta \theta { = }i\left( {{D_\mu }{\bm\phi }} \right)
\end{equation}
With this definition, the gauge covariant derivative commutes with the gauge transformation.
\begin{equation}
\label{2gaugeComm}
\left[ {\tfrac{\delta }{{\delta \theta }},{D_\mu }} \right]{\bm\phi  = }0
\end{equation}
Postpone investigating the construction of the gauge covariant derivative until later.

Now investigate the gauge transformation properties of the Lagrangian. Apply the chain rule to the Lagrangian as in Eq.~(\ref{chainR}), but with respect to the infinitesimal gauge transformation..
\begin{equation}
\label{gvar}
\frac{{\delta L}}{{\delta \theta }} = \frac{{\partial L}}{{\partial {\bm\phi }}}\frac{{\delta {\bm\phi }}}{{\delta \theta }}{ + }\frac{{\partial L}}{{\partial {D_\mu }{\bm\phi }}}\frac{{\delta \left( {{D_\mu }{\bm\phi }} \right)}}{{\delta \theta }} + \frac{{\partial L}}{{\partial {g^{\lambda \kappa }}}}\frac{{\delta \left( {{g^{\lambda \kappa }}} \right)}}{{\delta \theta }}\end{equation} Substitute the infinitesimal gauge transformation of the field, and note that the metric is defined to be invariant with respect to gauge transformations,
\begin{equation}
\label{metricVar}
\delta \left( {{g^{\lambda \kappa }}\left( x \right)} \right)/\delta {\theta ^a} = 0
\end{equation}
with the result,
\begin{equation}
\label{metricVar}
\frac{{\delta L}}{{\delta \theta }} = \frac{{\partial L}}{{\partial {\bm\phi }}}i{\bm\phi  + }\frac{{\partial L}}{{\partial {D_\mu }{\bm\phi }}}i\left( {{D_\mu }{\bm\phi }} \right)
\end{equation}
This is the response of the Lagrangian to an infinitesimal gauge transformation defined by Eqs.~(\ref{gaugeTransDeriv}) and (\ref{dgtrans}).

Apply Leibnitz rule to expose the Euler-Lagrange equation in the square bracket below. 
\begin{equation}
\label{LVarDeriv3}
\frac{{\delta L}}{{\delta \theta }} = \left[ {\frac{{\partial L}}{{\partial {\bm\phi }}}{ - }{D_\mu }\left( {\frac{{\partial L}}{{\partial {D_\mu }{\bm\phi }}}} \right)} \right]i{\bm\phi  + }{D_\mu }\left( {i\frac{{\partial L}}{{\partial {D_\mu }{\bm\phi }}}{\bm\phi }} \right)
\end{equation}
Substitute the row matrix ${\bm\Lambda }$ for the Euler-Lagrange expression in Eq.~(\ref{ELequ}), and \textit{define} the current as
\begin{equation}
\label{current}
j^\mu  = i\frac{{\partial L}}{{\partial {D_\mu }{\bm\phi }}}{\bm\phi }
\end{equation}
then substitute these quantities into Eq.~(\ref{LVarDeriv3}).
\begin{equation}
\label{LVarDeriv4}
\delta L/\delta \theta  = \bm\Lambda i{\bm\phi  + }{D_\mu }{j^\mu }
\end{equation}
A more interesting form of this equation appears when solved for the divergence of the current which is true even for Lagrangians without a particular symmetry \cite{Aitchison1}.
\begin{equation}
\label{divCurr}
\boxed{{D_\mu }{j^\mu } = \delta L/\delta \theta  - i\bm\Lambda {\bm\phi }}
\end{equation}
This relation shows that \textit{vanishing} gauge current divergence,
\begin{equation}
\label{currcons}
{D_\mu }j^\mu  = 0
\end{equation}
may be derived from 1) satisfaction of the Euler-Lagrange Eq.~(\ref{ELequ}), and 2) a \textit{symmetry} of the Lagrangian defined as,
\begin{equation}
\label{Lagrangiansymmetry}
\delta L/\delta \theta = 0
\end{equation}
with respect to 3) an infinitesimal gauge transformation. The current is defined in Eq.~(\ref{current}) whether or not the gauge transformation is a symmetry of the Lagrangian. In electromagnetics, the vanishing divergence of the current in  Eq.~(\ref{currcons}) is a continuity equation for electric current\cite{LandauContinuity,ContinuityEq}

\section{Construct the gauge covariant derivative}
Now define the form of the gauge covariant derivative. The above derivations require the following properties for the gauge covariant derivative,
\begin{itemize} \item is generally noncommuting, \item commutes when there are no external forces, \item is gauge covariant, \end{itemize}
 The gauge covariant derivative is defined the usual way by including the electromagnetic potential by the method of \textit{minimum coupling}\cite{Aitchison}~~\cite{FelsagerMinCouple}. For our purposes, minimum coupling means that the externally defined electromagnetic potential  gauge field ${A_\mu }\left( x \right)$ appears only in the covariant derivative, and no place else in the Lagrangian, justifying the functional dependency in the Lagrangian in Eq.~(\ref{Lagrangian}).

It is important to note that the form of the gauge covariant derivative depends upon the transformation properties of the operand. For example, it is an ordinary partial derivative when acting upon a coordinate and gauge invariant object.

Because the electromagnetic potential field does not have an equation of motion specified by the Lagrangian, Eq.~(\ref{Lagrangian}), it is therefore an externally applied field. The equation of motion for the wave function derived from the Euler-Lagrange equation includes the effect of external fields through the use of the gauge covariant derivative.

Following the deep analogy with coordinate covariant derivatives\cite{deepAnal}~~\cite{tensorDeriv}, define the gauge covariant derivative of the wave function ${\bm\phi }$ as\cite{theGCD}~~\cite{RamondGCD}
\begin{equation}
\label{gaugeCov}
{D_\mu }{\bm\phi } = {\partial _\mu }{\bm\phi  - }iA_\mu {\bm\phi }
\end{equation}
where the electromagnetic potential field $A_\mu \left( x \right)$ is now introduced. Absorb the electromagnetic coupling constant $e$  into the gauge potential $A_\mu$.

The gauge covariant derivative of the wave function must transform just as the wave function as indicated in Eq.~(\ref{dgtrans}).

Compute the commutator by substituting the gauge covariant derivative, Eq.~(\ref{gaugeCov}), revealing the connection to the electromagnetic field,
\begin{equation}
\label{gcdComm}
\left[ {{D_\mu },{D_\nu }} \right]{\bm\phi } = iF_{\mu \nu }{\bm\phi }
\end{equation}
where the definition of the electromagnetic \textit{field strength tensor} is
\begin{equation}
\label{gaugeFieldTensor}
{F_{\mu \nu }} =  - {\partial _\mu }{A_\nu } + {\partial _\nu }{A_\mu }
\end{equation}
The commutation of the derivative appears in Eq.~(\ref{theIdent}). The field strength tensor will allow us to identify the Lorentz force in the next section. The familiar vector component notation for the electromagnetic field tensor is\cite{Carmelip106}
\begin{equation}
\label{EMFieldTensor}
{F_{\mu \nu }} = \left( {\begin{array}{*{20}{c}}   0&{ - {E_x}}&{ - {E_y}}&{ - {E_z}} \\    {{E_x}}&0&{{H_z}}&{ - {H_y}} \\    {{E_y}}&{ - {H_z}}&0&{{H_x}} \\    {{E_z}}&{{H_y}}&{ - {H_x}}&0  \end{array}} \right)
\end{equation}

\section{Verify the transformation properties of the gauge covariant derivative}
The electromagnetic potential ${A_\mu }$ must respond to an infinitesimal gauge transformation in such a way as to effect the transformation properties of the gauge covariant derivative. Begin with the defined transformation properties for the gauge covariant derivative,  Eq.~(\ref{dgtrans}), then substitute the definition of the derivative, Eq.~(\ref{gaugeCov}).
\begin{equation}
\label{subDefgcd}
\delta \left( {{\partial _\mu }{\bm\phi  - }i{A_\mu }{\bm\phi }} \right){ = }\left( {i\delta \theta } \right)\left( {{\partial _\mu }{\bm\phi  - }i{A_\mu }{\bm\phi }} \right)
\end{equation}
Assume that the infinitesimal gauge transformation commutes with the partial derivative, Eq.~(\ref{2gaugeComm}). Only the transformation properties of the electromagnetic  potentials is unknown in the following.
\begin{equation}
\label{unkGPTrans}
{\partial _\mu }\delta {\bm\phi  - }i\left( {\delta {A_\mu }} \right){\bm\phi  - }i{A_\mu }\delta {\bm\phi  = }\left( {i\delta \theta } \right){\partial _\mu }{\bm\phi  + }\left( {\delta \theta } \right){A_\mu }{\bm\phi }
\end{equation}
Substitute the transformation of the wave function, Eq.~(\ref{gtrans}), on the left hand side then solve for the term with the unknown transformation of the gauge potentials.
\begin{equation}
\label{anyGen}
\delta {A_\mu }{ = }{\partial _\mu }\left( {\delta \theta } \right)
\end{equation}
This is the required gauge transformation property of the electromagnetic potential which verifies the transformation properties of the gauge covariant derivative, Eq.~(\ref{dgtrans}).

\section{The Lorentz Force}
We now have the definitions available to show that the electromagnetic field tensor acts on the current to subject the system to the Lorentz force.  Substitute the result of the commutation Eq.~(\ref{gcdComm}) into the differential relation Eq.~(\ref{theIdent}),
\begin{equation}
\label{divEMT}
{D_\nu }\left( {{T_\mu }^\nu } \right) = i\frac{{\partial L}}{{\partial {D_\nu }{\bm\phi }}}{F_{\mu \nu }}{\bm\phi } + \bm\Lambda {D_\mu }{\bm\phi }
\end{equation}
Substitute the current defined in Eq.~(\ref{current}).
\begin{equation}
\label{divEMTfinal}
\boxed{{D_\nu }\left( {{T_\mu }^\nu } \right) = F_{\mu \nu }{j^\nu } + \bm\Lambda {D_\mu }{\bm\phi }}
\end{equation}
This is the final form of the three-part relationship we seek between the divergence of the energy momentum tensor, the Lorentz force, and the Euler-Lagrange equation. Deserving more emphasis in the literature, we refer to the ``Lorentz force \textit{relation}'', which emphasizes the central importance of the Lorentz force.\cite{LandauR21,CarmeliLorentzForce} In this form, it becomes clear that the external field ${F_{\mu \nu }}$ acts on the current ${j^\mu }$, and so exchanging energy with the system as determined by the divergence of the energy momentum tensor.

The electromagnetic field tensor ${F_{\mu \nu }}$  is externally defined and so has no equation of motion. Similarly, the metric has no equation of motion.

The Lorentz force relation does \textit{not} require that the divergence of the current vanish since the current is defined by the infinitesimal gauge transformation, whether or not it is a symmetry of the Lagrangian.

Divergence of the energy momentum tensor equals a Lorentz force\cite{BarutP139} in the presence of an external gauge field acting on a charged current
\begin{equation}
\label{theLorentzForce}
{D_\nu }\left( {{T_\mu }^\nu } \right) = F_{\mu \nu }^a{j^\nu }
\end{equation}
-- a powerful connection made with a local procedure applied to the Lagrangian.

\section{Example: Lagrangian for the electromagnetic field }
The Lagrangian for the electromagnetic field\cite{AitchisonP26}
\begin{equation}
\label{EMLagrangian}
{L_{elect}} = {F^{\kappa \eta }}{F_{\kappa \eta }}
\end{equation}
also satisfies a relation \textit{similar} to the Lorentz force relation, Eq.~(\ref{divEMTfinal}). There are only slight differences in the development. We will discover the equations of motion as part of the differential relation satisfied by the Lagrangian, as before.

Here we pause and define the \textit{tensor dual} which is required for the Lorentz force relation. The definition of the \textit{ordinary} tensor dual\cite{CarmeliDualDef} is,
\begin{equation}
\label{dualDef}
\begin{gathered}   {{\tilde F}^{\mu \nu }} = \tfrac{1}{2}\tfrac{1}{{\sqrt { - g} }}{\varepsilon ^{\mu \nu \rho \sigma }}{F_{\rho \sigma }} \hfill \\   {{\tilde F}_{\rho \sigma }} =  - \tfrac{1}{2}\sqrt { - g} {\varepsilon _{\rho \sigma \alpha \beta }}{F^{\alpha \beta }} \hfill \\  \end{gathered} 
\end{equation}
where antisymmetric permutation tensors are defined,
\begin{equation}
\label{antisymDef}
{\varepsilon ^{0123}} = 1
\end{equation}
and the determinant of the space-time metric is $g$. Application of the dual twice, results in the negative of the original antisymmetric tensor for the space-time metric which has a negative determinant.
\begin{equation}
\label{antisymDef}
- {F_{\mu \nu }} = \tilde {\tilde{F_{\mu \nu }}}
\end{equation}

Now returning to our calculation, apply the covariant derivative to the Lagrangian.
\begin{equation}
\label{covDerivLag}
{\nabla _\mu }{L_{elect}} = 2{\nabla _\mu }\left( {{F_{\kappa \eta }}} \right){F^{\kappa \eta }}
\end{equation}
Substitute the useful relation\cite{SchwingRel} where $\tilde F$ indicates the tensor dual of $F$, and $G$ is an arbitrary antisymmetric second order tensor.
\begin{equation}
\label{SchwingRelation}
\left( {{\nabla ^\eta }{F^{\kappa \mu }}} \right){G_{\kappa \eta }} - \tfrac{1}{2}\left( {{\nabla ^\mu }{F^{\kappa \eta }}} \right){G_{\kappa \eta }} = {\tilde G^{\mu \lambda }}\left( {{\nabla _\nu }\tilde F_{\;\;\lambda }^\nu } \right)
\end{equation}
or, moving indices, and apply to our case where $G=F$
\begin{equation}
\label{electRel1}
\left( {{\nabla _\mu }{F_{\kappa \eta }}} \right){F^{\kappa \eta }} = 2\left( {{\nabla _\eta }{F_{\kappa \mu }}} \right){F^{\kappa \eta }} - 2{\tilde F_{\mu \lambda }}\left( {{\nabla _\nu }\tilde F_{\;\;\;\;}^{\nu \lambda }} \right)
\end{equation}
then substitute this relation into Eq.~(\ref{covDerivLag}).
\begin{equation}
\label{electRel2}
{\nabla _\mu }{L_{elect}} = 4{\nabla _\eta }\left( {{F_{\kappa \mu }}} \right){F^{\kappa \eta }} - 4{\tilde F_{\mu \lambda }}{\nabla _\nu }\left( {{{\tilde F}^{\nu \lambda }}} \right)
\end{equation}
In first term on the right, pull the second field tensor into the derivative as done several times prior, then subtract the additional term arising from the product rule.
\begin{equation}
\label{electRel3}
\begin{gathered}
  {\nabla _\mu }{L_{elect}} = 4{\nabla _\eta }\left( {{F_{\kappa \mu }}{F^{\kappa \eta }}} \right) \\ 
   - 4{F_{\kappa \mu }}{\nabla _\eta }\left( {{F^{\kappa \eta }}} \right) - 4{{\tilde F}_{\mu \lambda }}{\nabla _\nu }\left( {{{\tilde F}^{\nu \lambda }}} \right) \\ 
\end{gathered} 
\end{equation}
Substitute the Kronecker delta tensor, and the definition of the Lagrangian in Eq.~\ref{EMLagrangian}.
\begin{equation}
\label{electRel4}
\begin{gathered}
  {\nabla _\eta }\left[ {\tfrac{1}{4}\delta _\mu ^\eta {L_{elect}} - {F_{\kappa \mu }}{F^{\kappa \eta }}} \right] =  - {F_{\kappa \mu }}{\nabla _\eta }\left( {{F^{\kappa \eta }}} \right) \\ 
   - {{\tilde F}_{\mu \lambda }}{\nabla _\nu }\left( {{{\tilde F}^{\nu \lambda }}} \right) \\ 
\end{gathered}
\end{equation}
Identify the energy momentum tensor for the electromagnetic field,
\begin{equation}
\label{electRel5}
\boxed{T_{\;\;\mu \,\left( {em} \right)}^\eta  = \tfrac{1}{4}\delta _\mu ^\eta {F^{\kappa \gamma }}{F_{\kappa \gamma }} - {F_{\kappa \mu }}{F^{\kappa \eta }}}
\end{equation}
and define the \textit{magnetoelectric} currents
\begin{equation}
\label{MECurrents}
\begin{gathered}
  j_e^\kappa  = {\nabla _\eta }\left( {{F^{\eta \kappa }}} \right) \hfill \\
  j_m^\kappa  = {\nabla _\nu }\left( {{{\tilde F}^{\nu \lambda }}} \right) \hfill \\ 
\end{gathered} 
\end{equation}
Substitute these into Eq.~\ref{electRel4} to obtain the \textit{generalized} Lorentz force equation analogous to Eq.~\ref{divEMTfinal}.
\begin{equation}
\label{generalizedLorentzForce}
{\nabla _\nu }T_{\;\;\mu \,\left( {em} \right)}^\nu  = {F_{\nu \mu }}j_e^\nu  + {\tilde F_{\nu \mu }}j_m^\nu 
\end{equation}
In this electric/magnetic symmetrized form, $j_m^\nu $ is regarded as a current of magnetic charge which is set to zero to obtain the usual Maxwell's equations for electromagnetics.
\begin{equation}
\label{MaxEqu}
\begin{gathered}
  j_e^\kappa  = {\nabla _\eta }\left( {{F^{\eta \kappa }}} \right) \hfill \\
  0 = j_m^\kappa  = {\nabla _\nu }\left( {{{\tilde F}^{\nu \lambda }}} \right) \hfill \\ 
\end{gathered}  
\end{equation}
so that the Lorentz force for the electromagnetic field is,\cite{JacksonP611}
\begin{equation}
\label{totalSystem}
{\nabla _\nu }T_{\;\;\mu \,\left( {em} \right)}^\nu  = {F_{\nu \mu }}j_e^\nu 
\end{equation}
The charged currents giving rise to the electromagnetic field in Maxwell's equations is not restricted to being identical to the charged current associated with the wave function. Both have vanishing divergence, but simply summing the Lorentz forces in Eqs.~\ref{generalizedLorentzForce} and \ref{KGLorentzForce} results in,
\begin{equation}
\label{totalSystem}
{\nabla _\nu }T_{\;\;\mu \,\left( {em} \right)}^\nu  + {\nabla _\nu }T_{\mu }^{\;\;\;\nu } = {F_{\nu \mu }}\left( {j_e^\nu  + {j^\nu }} \right)
\end{equation}

The currents summing to zero would satisfy the additional imposition that the net forces sum to zero for an isolated system. There are many more complexities which could be added to this simple model.

\section{The necessity of contravariance}
A typical goal in constructing a Lagrangian is to require gauge invariance of the Lagrangian, so that the gauge current will have a ``vanishing divergence'', Eq.~(\ref{currcons}), which is current continuity in the case of electromagnetic currents. The transformation property of \textit{contravariance} is defined so that an inner product between a covariant object and a contravariant object is a gauge invariant object, hence a candidate for a term in the Lagrangian. The covariant property of the wave function is defined in Eq.~(\ref{gtrans}). We have already encountered contravariance in the Euler-Lagrange equations,  Eq.~(\ref{ELequ}), which are contravariant, when the Lagrangian is invariant.

By definition, a \textit{contravariant} and otherwise arbitrary wave function $\bm\psi$ (a row matrix) in a matrix inner product with the covariant wave function $\bm\phi$ (a column matrix) is gauge \textit{invariant},
\begin{equation}
\label{invarInnerProd}
\delta \left( {\bm\psi {\bm\phi }} \right) = 0
\end{equation}
Apply the product rule for derivatives to find that the following transformation of the contravariant wave function is required to achieve invariance.
\begin{equation}
\label{contravDef}
\delta \bm\psi /\delta \theta { = } - i\bm\psi
\end{equation}
This transformation defines the contravariant transformation.

Define the derivative applied to an arbitrary contravariant wave function $\bm\psi$ as \begin{equation}
\label{contravDefDeriv}
{D_\mu }\bm\psi  = {\partial _\mu }\bm\psi  + i{A_\mu }\bm\psi
\end{equation} so that the gauge covariant derivative reduces to the ordinary partial derivative for the invariant object $\left( {\bm\psi {\bm\phi }} \right)$, \begin{equation}
\label{derivOfInvar}
{D_\mu }\left( {\bm\psi {\bm\phi }} \right) = {\partial _\mu }\left( {\bm\psi {\bm\phi }} \right)
\end{equation}

The commutation of the gauge covariant derivative as applied to the contravariant wave function becomes, \begin{equation}
\label{commutationContravar}
\left[ {{D_\mu },{D_\nu }} \right]\bm\psi  =  - i{F_{\mu \nu }}\bm\psi 
\end{equation} With these extensions, the derivative and gauge transformation in Eqs.~(\ref{contravDefDeriv}) and (\ref{contravDef}) can now be consistently applied to contravariant expressions.

\section{The contravariant wave function}
The construction of an invariant Lagrangian requires inner products as in Eq.~(\ref{invarInnerProd}). How to construct a contravariant wave function without introducing fields in addition to the components of the column matrix wave function ${\bm\phi }$? We can take a hint from Dirac spinors and construct a row matrix wave function ${\bar {\bm\phi} }$ with the required \textit{contravariant} gauge transformation properties, Eq.~(\ref{contravDef}). We assume the components of the wave function are complex numbers, so that the matrix conjugate transpose is,
\begin{equation}
\label{defAdjoint}
{{\bm\phi }^\dag } = {\left( {{{\bm\phi }^T}} \right)^*}
\end{equation}

We \textit{define} the contravariant version (or dual, or adjoint) of a wave function ${\bm\phi }$ as,\cite{MessAdjointWF}
\begin{equation}
\label{defContravWFAdjoint}
{\bar {\bm\phi}  = }{{\bm\phi }^\dag }
\end{equation}
which has exactly the required contravariant transformation properties, Eq.~(\ref{contravDef})
\begin{equation}
\label{defContravTransf}
\begin{gathered}   \delta {\bar {\bm\phi} }/\delta \theta { = } - i{\bar {\bm\phi} } \hfill \\   \delta {{\bm\phi }^\dag }/\delta \theta { = } - i{{\bm\phi }^\dag } \hfill \\  \end{gathered}
\end{equation}
The definition in Eq.~(\ref{defContravWFAdjoint}) also assures a \textit{positive-definite} norm for the wave function using a simple matrix inner product. Use a large parenthesis with comma to indicate an inner product.\cite{SchiekInnerProd}
\begin{equation}
\label{innerProd}
\left( {{\bm\phi ,\bm\phi }} \right) \equiv {\bar {\bm\phi} {\bm\phi}  = }{{\bm\phi }^\dag }{\bm\phi } \in \mathbb{R}
\end{equation} 

The contravariant derivative applied to a contravariant wave is,
\begin{equation}
\label{defContravWFDeriv}
{D_\mu }{{\bm\phi }^\dag } = {\left( {{D_\mu }{\bm\phi }} \right)^\dag } = {\partial _\mu }{{\bm\phi }^\dag } + i{A_\mu }{{\bm\phi }^\dag }
\end{equation}
assuming that the conjugate transpose operation commutes with the derivative.

\section{Conclusion}
A simple differential relation for the Lagrangian Eq.~(\ref{theIdent}) becomes a relation between physical entities, Eq.~(\ref{divEMTfinal}), with the identification of the  Lorentz force acting on a charged current. The Lorentz force provides the connection between the changing energy of a system, and the equations of motion. Intuitively, the equations of motion must define energy flow between the system and an external field, and Eq.(\ref{divEMTfinal}) quantifies that intuition, and without invoking the principle of least action.

The local approach provides sufficient motivation for the Euler-Lagrange equations as the equation of motion defined by a Lagrangian. It is interesting that the local approach allows consideration of ``off-shell'' equations of motion, and is independent of space or time symmetries, or the principle of least action.
 
 The definition of the charged current arises from the definition of the gauge transformation, whether or not it is a symmetry of the Lagrangian. The charged current is required in the definition of the Lorentz force. The charged current is acted upon by the external field as the Lorentz force.

The problem of constructing a gauge invariant Lagrangian requires covariant and contravariant objects.

A specific quadratic Lagrangian for the Klein-Gordon wave equation of motion, demonstrates how the Euler-Lagrange equation yields the wave equation.

The Lagrangian for the electromagnetic field provides a similar Lorentz force connection to charged currents.

 We hope that the local approach presented here leads to a deeper understanding of Lagrangian mechanics for the student.

\begin{acknowledgments}

I thank Jacques Rutschmann for providing just the right amount of skepticism and encouragement.

\end{acknowledgments}

\appendix   % Omit the * if there's more than one appendix.

\section{Example: the wave function in an external electromagnetic field}
Examine a specific Lagrangian for a wave function consisting of a column matrix of complex scalars, represented as $\bm\phi$, which will yield a similar wave equation as the relativistic Klein-Gordon wave equation.\cite{SakuraiKGEq} Investigate the  scalar and gauge invariant Lagrangian for this system.
\begin{equation}
\label{KGLagr}
L\left( {{\bm\phi },{D_\mu }{\bm\phi ,}{g_{\mu \nu }}} \right) = {D^\lambda }{{\bm\phi }^\dag }{D_\lambda }{\bm\phi } - {M^2}{{\bm\phi }^\dag }{\bm\phi }
\end{equation}
As in Eq.~(\ref{chainR}), apply the gauge covariant derivative to the Lagrangian. We follow the same steps leading to Eq.~(\ref{theIdent}), but twice, for the wave function and its adjoint.
\begin{equation}
\label{KGLagrDeriv}
\begin{gathered}   {D_\mu }L = {D_\mu }\left( {{D^\lambda }{{\bm\phi }^\dag }} \right){D_\lambda }{\bm\phi } + {D^\lambda }{{\bm\phi }^\dag }{D_\mu }\left( {{D_\lambda }{\bm\phi }} \right) \\     - {M^2}{D_\mu }\left( {{{\bm\phi }^\dag }} \right){\bm\phi } - {M^2}{{\bm\phi }^\dag }{D_\mu }{\bm\phi } \\  \end{gathered} 
\end{equation}
Add and subtract the second order derivative, but with swapped indices. Use a commutator bracket.
\begin{equation}
\label{KGLagrComm}
\begin{gathered}   {D_\mu }L = {D^\lambda }\left( {{D_\mu }{\bm\phi ^\dag}} \right){D_\lambda }\bm\phi  + {D^\lambda }{\bm\phi ^\dag}{D_\lambda }\left( {{D_\mu }\bm\phi } \right) \\     + \left( {\left[ {{D_\mu },{D^\lambda }} \right]{\bm\phi ^\dag}} \right){D_\lambda }\bm\phi  + {D^\lambda }{\bm\phi ^\dag}\left( {\left[ {{D_\mu },{D_\lambda }} \right]\phi } \right) \\     - {M^2}{D_\mu }\left( {{\bm\phi ^\dag}} \right)\phi  - {M^2}{\bm\phi ^\dag}{D_\mu }\bm\phi  \\  \end{gathered} 
\end{equation}
Apply the product rule, and rearrange to expose the Euler-Lagrange equations of motion which are the coefficient of the derivative of the field $\left( {{D_\mu }{\bm\phi }} \right)$ and its gauge contravariant derivative $\left( {{D_\mu }{{\bm\phi }^\dag }} \right)$.
\begin{equation}
\label{KGLagrEuler}
\begin{gathered}   {D_\mu }L = {D^\lambda }\left( {{D_\mu }{{\bm\phi }^\dag }{D_\lambda }{\bm\phi }} \right) - \left( {{D_\mu }{{\bm\phi }^\dag }} \right)\left( {{D^\lambda }{D_\lambda }{\bm\phi } + {M^2}{\bm\phi }} \right) \\     + {D_\lambda }\left( {{D^\lambda }{{\bm\phi }^\dag }{D_\mu }{\bm\phi }} \right) - \left( {{D_\lambda }{D^\lambda }{{\bm\phi }^\dag } + {M^2}{{\bm\phi }^\dag }} \right)\left( {{D_\mu }{\bm\phi }} \right) \\     + \left( {\left[ {{D_\mu },{D^\lambda }} \right]{{\bm\phi }^\dag }} \right){D_\lambda }{\bm\phi } + {D^\lambda }{{\bm\phi }^\dag }\left( {\left[ {{D_\mu },{D_\lambda }} \right]{\bm\phi }} \right) \\  \end{gathered} 
\end{equation}
Identify the Euler-Lagrange equation, and its conjugate transpose as coefficients of the covariant derivative of the wave function which is the only first order derivative of the wave function with a free index $\mu$. Once the expression is brought into this form, identification of the Euler-Lagrange equation $\bm\Lambda$ is unique.
\begin{equation}
\label{KGWave}
\begin{gathered}   {D^\lambda }{D_\lambda }\bm\phi  + {M^2}\bm\phi  = 0 \hfill \\   {D_\lambda }{D^\lambda }{\bm\phi ^\dag} + {M^2}{\bm\phi ^\dag} = 0 \hfill \\  \end{gathered} 
\end{equation}
Substitute, and collect terms under the derivative. \begin{equation}
\label{collectTermsKG}
\begin{gathered}   {D_\mu }L = {D_\lambda }\left( {{D_\mu }{\bm\phi ^\dag}{D^\lambda }\bm\phi  + {D^\lambda }{\bm\phi ^\dag} {D_\mu }\bm\phi } \right) \\     + \left( {\left[ {{D_\mu },{D^\lambda }} \right]{\bm\phi ^\dag}} \right){D_\lambda }\bm\phi  + {D^\lambda }{\bm\phi ^\dag}\left( {\left[ {{D_\mu },{D_\lambda }} \right]\phi } \right) \\  \end{gathered} 
\end{equation} Insert the Kronecker delta, then move terms to the left hand side. Rename dummy indices.
\begin{equation}
\label{KGInsertKronecker}
\begin{gathered}    - {D_\nu }\left[ {{D_\mu }{\bm\phi ^\dag}{D^\nu }\bm\phi  + {D^\nu }{\bm\phi ^\dag} {D_\mu }\bm\phi  - \delta _\mu ^\nu L} \right] =  \hfill \\   \left( {\left[ {{D_\mu },{D^\lambda }} \right]{\bm\phi ^\dag}} \right){D_\lambda }\bm\phi  + {D^\lambda }{\bm\phi ^\dag}\left( {\left[ {{D_\mu },{D_\lambda }} \right]\phi } \right) \hfill \\  \end{gathered}
\end{equation}
The energy-momentum tensor ${T_{KG\mu \nu }}$ for wave function appears inside the large square bracket.
\begin{equation}
\label{KGEMT}
{T_{KG\mu \nu }} = {D_\mu }{\bm\phi ^\dag}{D_\nu }\bm\phi  + {D_\nu }{\bm\phi ^\dag} {D_\mu }\bm\phi  - {g_{\mu \nu }}L
\end{equation}
Substitute the commutator results, Eq.~(\ref{gcdComm}) and (\ref{commutationContravar}) into Eq.~(\ref{KGInsertKronecker}).\cite{MessCommutation}
\begin{equation}
\label{KGsubComm}
{D_\nu }T_{KG\mu }^{\;\;\;\nu } = ie{F_{\mu \nu }}\left( {{{\bm\phi }^\dag }{D^\nu }{\bm\phi } - \left( {{D^\nu }{{\bm\phi }^\dag }} \right){\bm\phi }} \right)
\end{equation}
Substitute the definition of the charged current,
\begin{equation}
\label{KGcurrent}
{j^\nu } = ie\left( {{{\bm\phi }^\dag }{D^\nu }{\bm\phi } - \left( {{D^\nu }{{\bm\phi }^\dag }} \right){\bm\phi }} \right)
\end{equation}
then recognize the Lorentz force on the right hand side of the following equation.\cite{LandauRefLForce}
\begin{equation}
\label{KGLorentzForce}
{\nabla _\nu }T_{KG\mu }^{\;\;\;\nu } = {F_{\mu \nu }}{j^\nu }
\end{equation}
The divergence of the canonical energy-momentum tensor is equal to the Lorentz force as required in Eq.~(\ref{divEMTfinal}). Identification of the Lorentz force is one of the rewards for assuming satisfaction of the Euler-Lagrange equation for this wave function, Eq.~(\ref{KGWave}).

The column matrix form of the wave function may be made explicit by indexing the wave function components.
\begin{equation}
\label{indexedWF}
{\bm\phi  = }\left( {{\phi _i}} \right)
\end{equation}
then substituting into the Lagrangian for the scalar field, Eq.~(\ref{KGLagr}), find that the Lagrangian becomes a sum of individual Lagrangians.
\begin{equation}
\label{LagrSumI}
L = \sum\limits_i {{L_i}} 
\end{equation}
where
\begin{equation}
\label{LagrI}
{L_i} = {D^\lambda }\phi _i^\dag {D_\lambda }{\phi _i} - {M^2}\phi _i^\dag {\phi _i}
\end{equation}
Each component satisfies the Euler-Lagrange equation.
\begin{equation}
\label{ELEquI}
\Lambda \left( {{L_i}} \right) = {D^\lambda }{D_\lambda }{\phi _i} + {M^2}{\phi _i}{ = 0}
\end{equation}
Similarly, the quadratic form of the current ${j^\lambda }$ in Eq.~(\ref{KGcurrent}), splits into a sum of components.
\begin{equation}
\label{KGCurrSumI}
{j^\nu } = \sum\limits_i {j_i^\nu } 
\end{equation}
where
\begin{equation}
\label{KGCurrI}
j_i^\nu  = ie\left( {{\phi }_i^\dag {D^\nu }{{\phi }_i} - \left( {{D^\nu }{\phi }_i^\dag } \right){{\phi }_i}} \right) 
\end{equation}
and again for the canonical energy momentum tensor,
\begin{equation}
\label{KGEMTSum}
T_{KG\mu }^{\;\;\;\nu } = \sum\limits_i {T_{iKG\mu }^{\;\;\;\nu }} 
\end{equation}

We see that each complex scalar component of the column matrix of the field ${\bm\phi }$ ''feels'' the same external electromagnetic field, and evolves independently of the other components of the field, without interaction between the scalar field components.
\begin{equation}
\label{KGEMTI}
{\nabla _\nu }T_{iKG\mu }^{\;\;\;\nu } = {F_{\mu \nu }}j_i^\nu
\end{equation}
An interesting task would be to extend the model, including interactions between the field components.

\section{Example: define the charge current}

The identification of the Lorentz force in Eq.~(\ref{KGLorentzForce}) depends upon the definition of the current. The charged current is defined in terms of the infinitesimal gauge transformation, Eq.~(\ref{current}), but only if we are not so quick to cancel terms. Carry through the variations, starting with the chain rule, just as in Eq.~(\ref{gvar}).
\begin{equation}
\label{KGSym1}
\delta L = \delta {D^\lambda }{\bm\phi ^\dag}{D_\lambda }\bm\phi  + {D^\lambda }{\bm\phi ^\dag}\delta {D_\lambda }\bm\phi  - {M^2}\delta {\bm\phi ^\dag}\bm\phi  - {M^2}{\bm\phi ^\dag}\delta \phi 
\end{equation}
Upon applying the infinitesimal gauge transformation to this expression, it is immediately apparent that the Lagrangian is invariant, hence the variation vanishes. Substitute the variations, but do not cancel terms.
\begin{equation}
\label{KGSym3}
\frac{{\delta L}}{{ie\delta \theta }} = 0 =  - {D^\lambda }{\bm\phi ^\dag}{D_\lambda }\bm\phi  + {D^\lambda }{\bm\phi ^\dag}{D_\lambda }\bm\phi  + {M^2}{\bm\phi ^\dag}\bm\phi  - {M^2}{\bm\phi ^\dag}\bm\phi 
\end{equation}
It might seem strange working with terms which sum to zero, but that is exactly the step in Eq.~(\ref{LVarDeriv3}). In each of the two terms containing derivative, extract one of the derivatives to create a divergence, then subtract the extra term required by the product rule. Rearrange to expose the Euler-Lagrange wave equations, Eq.~(\ref{KGWave}).
\begin{equation}
\label{KGSym5}
\begin{gathered}   \frac{{\delta L}}{{ie\delta \theta }} = 0 =  - {D^\lambda }\left( {{\bm\phi ^\dag}{D_\lambda }\bm\phi } \right) + {\bm\phi ^\dag}\left( {{D^\lambda }{D_\lambda }\bm\phi  + {{M}^2}\bm\phi } \right) \\     + {D_\lambda }\left( {\bm\phi {D^\lambda }{\bm\phi ^\dag}} \right) - \phi \left( {{D_\lambda }{D^\lambda }{\bm\phi ^\dag} + {{M}^2}{\bm\phi ^\dag}} \right) \\  \end{gathered}
\end{equation}
Assume the Euler-Lagrange wave equations are zero for the wave function. The remaining terms which are the divergence of the gauge invariant current.
\begin{equation}
\label{KGSym6}
\frac{{\delta L}}{{ie\delta \theta }} = 0 = {D^\lambda }\left( {\bm\phi {D_\lambda }{\bm\phi ^\dag} - {\bm\phi ^\dag}{D_\lambda }\bm\phi } \right)
\end{equation}
where the current is defined exactly as Eq.~(\ref{KGcurrent}), so that the divergence of the current vanishes, as a consequence of Eq.~(\ref{KGSym6}) the invariant Lagrangian and the Euler-Lagrange equations of motion.
\begin{equation}
\label{KGSym7}
\delta L/\delta \theta  =  - {\nabla _\lambda }{j^\lambda } = 0
\end{equation}

This example shows that the considerations for a general Lagrangian can be applied to specific Lagrangians. The Euler-Lagrange equations for a general Lagrangian in Eq.~(\ref{theIdent}) or Eq.~(\ref{divCurr}) can easily be identified using the same local procedure for a specific Lagrangian.

\end{document}